\documentclass[conference]{IEEEtran}
\usepackage[T1]{fontenc}
\usepackage[latin9]{inputenc}
\usepackage{color}
\usepackage{bm}
\usepackage{amsthm}
\usepackage{amsmath}
\usepackage{amssymb}
\usepackage{graphicx}

\makeatletter
\theoremstyle{plain}
\newtheorem{thm}{\protect\theoremname}

\usepackage{cite}
\usepackage{psfrag}
\usepackage{flushend}
\usepackage{url}
\IEEEoverridecommandlockouts

\makeatother

\providecommand{\theoremname}{Theorem}

\begin{document}

\title{Distributed Compressed Sensing for\\Sensor Networks with Packet
Erasures}

\author{\authorblockN{Christopher Lindberg, Alexandre Graell i Amat, Henk
Wymeersch}\authorblockA{Department of Signals and Systems, Chalmers
University of Technology, Gothenburg, Sweden,\\
Email: \{chrlin,alexandre.graell,henkw\}@chalmers.se}\thanks{This
research was supported, in part, by the European Research Council,
under Grant No. 258418 (COOPNET), and the Swedish Research Council,
\textcolor{black}{under} Grant No. 2011-5961.}}
\maketitle
\begin{abstract}
We study two approaches to distributed compressed sensing for in-network
data compression and signal reconstruction at a sink. Communication
to the sink is considered to be bandwidth-constrained due to the large
number of devices. By using distributed compressed sensing for compression
of the data in the network, the communication cost (bandwidth usage)
to the sink can be decreased at the expense of delay induced by the
local communication. We investigate the relation between cost and
delay given a certain reconstruction performance requirement when
using basis pursuit denoising for reconstruction. Moreover, we analyze
and compare the performance degradation due to erased packets sent
to the sink.
\end{abstract}

\section{Introduction}

Wireless sensor networks (WSNs) provide a tool to accurately monitor
physical phenomena over large areas \cite{akyildiz2002wireless}.
The WSN is usually considered to be energy-constrained and to comprise
up to several thousands of nodes. However, smart phones and other
\textcolor{black}{sensing} devices carrying powerful batteries have
become ubiquitous. This provides a possible platform for WSNs where
energy is not a scarce resource. Instead, the sheer number of sensors
puts a strain on the bandwidth available for communication between
the sensors and the sink. Consequently, the measurement data acquired
by the sensors needs to be compressed. Compression should be able
to operate under unreliable communication conditions and be scalable
in the number of sensors. Existing techniques, such as Slepian-Wolf
coding and transform coding (see \cite{duarte2012signal} and references
therein) require precise statistical models about the phenomena of
interest. Compressed sensing (CS) \cite{candes2004decode,candes2005stable,donoho2004comp},
on the other hand, alleviates the need for precise statistical models
and is scalable in the number of sensors \cite{huang2013applications}.

Prior work on CS in WSNs includes \cite{bajwa2006cws,haupt2008csnd,luo2009compressive,ozheng2011csdel,patterson2013distributed}.
In \cite{bajwa2006cws,haupt2008csnd}, CS is used for in-network compression,
but communication to the sink is done by analog phase-coherent transmissions.
This is not practical for WSNs operating in a cellular network since
all sensors need to be perfectly synchronized. In \cite{luo2009compressive}
and \cite{ozheng2011csdel}, CS is considered for networks with multi-hop
routing towards the sink. In addition, \cite{ozheng2011csdel} considers
the delay caused by a medium access control (MAC) protocol. The drawback
of multi-hop routing is the necessity to form a spanning tree, which
is impractical and prone to communication failures, especially when
the sensors are mobile. In \cite{patterson2013distributed}, no sink
is present, but the sensors use CS and consensus to perform distributed
signal reconstruction. However, the focus is on reconstruction performance
and the MAC delay is not studied.

In this paper, we consider distributed CS for a WSN with equispaced
sensors on a straight line. The sensors sense a physical phenomenon
in their local environment\textcolor{black}{, perform in-network compression,}
and transmit the (compressed) data to a common sink. We analyze the
tradeoff between communication cost towards the sink and MAC delay
from the inter-sensor communication. We consider two approaches that
rely on local processing between sensors, where only a subset of the
nodes communicate to the sink. The first approach performs local processing
by clustering of the sensors, while the other uses average consensus.
Additionally, we compare the robustness to packet erasures when transmission
to the sink is performed over a noisy (erasure) channel. Our contributions
are:
\begin{itemize}
\item Closed-form expressions for the upper bound on the reconstruction
error for basis pursuit denoising (BPDN), that guarantees stable reconstruction
for both approaches in the presence of packet erasures.
\item Closed-form expressions for the communication cost and the MAC delay
to meet a given performance requirement for the consensus approach.
\end{itemize}
\emph{Notation:} We use boldface lowercase letters $\boldsymbol{x}$
for column vectors, and boldface uppercase letters $\boldsymbol{X}$
for matrices. In particular, $\boldsymbol{I}_{M}$ denotes an $M\times M$
identity matrix, $\boldsymbol{1}$ is the all-one vector, and $\boldsymbol{0}$
is the all-zero vector. Sets are described by caligraphic letters
$\mathcal{X}$. The cardinality of a set $\mathcal{X}$ is denoted
by $|\mathcal{X}|$. The transpose of a vector is denoted by $[\cdot]^{\mathsf{T}}$.
Expectation of a random variable is denoted by $\mathbb{E}\{\cdot\}$,
and $\mathrm{Var}(\cdot)$ indicates the variance of a random variable
or covariance matrix of a random vector. The indicator function of
a set $\mathcal{X}$ is written as $\mathbb{I}_{\mathcal{X}}(\cdot)$.

\section{System Model\label{sec:SM}}

\begin{figure}
\includegraphics[width=0.9\columnwidth]{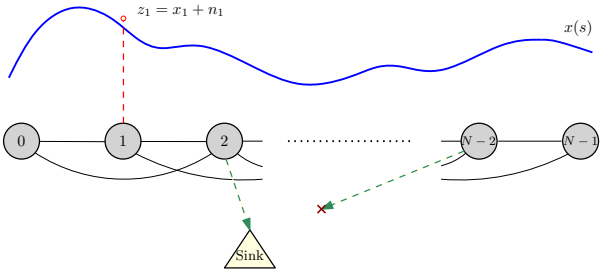}\caption{\label{fig:sysmod}Illustration of the system model. The grey circles
are the sensor nodes measuring the intensity of the signal $x(s)$.
They perform local processing using sensor-to-sensor communication
(black lines). Sensors $2$ and $N-2$ send packets to the sink, but
the packet from sensor $N-2$ is erased.}
\end{figure}

\subsection{\label{sub:commodel}Sensor and Network Model}

The system model is illustrated in Fig. \ref{fig:sysmod}. We consider
a one-dimensional network of $N$ nodes placed equally spaced on a
straight line. Without loss of generality we set the coordinate of
sensor $k$, $k=0,\dots,N-1$, to $s_{k}=k$. The sensors measure
the intensity of a real-valued signal $x(s)$ in their respective
spatial coordinates. The observation of sensor $k$ is
\begin{equation}
z_{k}=x_{k}+n_{k},
\end{equation}
where $x_{k}=x(k)$ and the $n_{k}$'s are spatially white Gaussian
noise samples with variance $\sigma_{n}^{2}$. The observations are
stacked in a vector $\boldsymbol{z}=[z_{0},\dots,z_{N-1}]^{\mathsf{T}}$.

Each node can communicate over a licensed spectrum with a base station
serving as a sink, or fusion center (FC), incurring a fixed cost (bandwidth
usage) $C$. The node-to-sink links are modeled as independent erasure
channels with packet erasure probability $p$.%
\footnote{The event that all packets sent to the sink are erased is not considered.%
} Communication from the nodes to the FC relies on orthogonal channels,
and thus incurs no delay.

The nodes can also exchange information locally with nearby nodes
using broadcasting over a shared (unlicensed) channel. To avoid packet
collisions the transmissions are scheduled using a spatial time division
multiple access (S-TDMA) protocol. Each node is allowed to transmit
only in an assigned time slot, which is reused by other nodes if they
are spatially far apart. Therefore, the local communication will incur
a delay $D$ (expressed in a number of TDMA slots), but is on the
other hand considered to be cost-free. We use a disc model with radius
$R$ to determine if two nodes are connected. For later use, we denote
by $\mathcal{G}=(\mathcal{V},\mathcal{E})$ the undirected graph describing
the network, where $\mathcal{V}$ is the set of nodes and $\mathcal{E}$
the set of edges connecting the nodes.

\subsection{Signal Model\label{sub:signalsensor}}

We consider a smooth, band-limited spatial signal $x(s)$, sampled
as $\boldsymbol{x}=[x_{0},\dots,x_{N-1}]^{\mathsf{T}}$ with energy
$E_{X}=\sum_{k=0}^{N-1}\left|x_{k}\right|^{2}$. Furthermore, we assume
that there exists a transformation $\boldsymbol{\theta}=\boldsymbol{Tx}$
such that $\boldsymbol{\theta}\in\mathbb{C}^{N\times1}$ is $K$-sparse,
i.e., $\boldsymbol{\theta}$ has $K\ll N$ nonzero elements. In our
case, the signal $x(s)$ is regarded as sparse in the spatial frequency
domain, owing to the smoothness of $x(s)$. Since nodes are equispaced,
we can use a discrete Fourier transform (DFT) matrix as $\boldsymbol{T}$,
with entries
\begin{equation}
T_{ml}=\frac{1}{\sqrt{N}}\exp\left(-j\frac{2\pi ml}{N}\right),
\end{equation}
for $m,l\in\{0,\dots,N-1\}$. The entries of $\boldsymbol{\theta}$
are then the sampled spatial frequencies of $x(s)$. We will denote
the average signal-to-noise ratio per sample $E_{X}/(N\sigma_{n}^{2})$
as $\mathsf{SNR}$.

\subsection{Goal}

Given the observations $\boldsymbol{z}$ and the system model outlined
above, the goal is to reconstruct $\boldsymbol{x}$ at the sink such
that a certain reconstruction error is guaranteed.

\section{Compressed Sensing Background\label{sec:csback}}

\subsection{Definition and Performance Measure\label{sub:defper}}

Let $\boldsymbol{x}$ and $\boldsymbol{Tx}=\boldsymbol{\theta}$ be
as described in Section \ref{sub:signalsensor}. Also, let $\boldsymbol{A}\in\mathbb{R}^{M\times N}$
be a measurement matrix where $M\ll N$, and define the compression
\begin{equation}
\boldsymbol{y}=\boldsymbol{Az}=\boldsymbol{Ax}+\boldsymbol{e}_{\mathrm{obs}},\label{eq:compress2}
\end{equation}
where $\boldsymbol{e}_{\mathrm{obs}}=\boldsymbol{An}$\textcolor{black}{,
and}\textcolor{blue}{{} }\textcolor{black}{$\bm{n}=\left[n_{0},\dots,n_{N-1}\right]^{\mathsf{T}}$.
}Since $M\ll N$, recovering $\boldsymbol{x}$ from $\boldsymbol{y}$
is an ill-posed problem, as there are infinitely many vectors $\tilde{\boldsymbol{x}}$
that satisfy $\boldsymbol{y}=\boldsymbol{A}\tilde{\boldsymbol{x}}$.
 However, we can exploit the knowledge about the sparsity of $\boldsymbol{x}$
in the transform domain. If $\boldsymbol{A}$ satisfies the restricted
isometry property (RIP) \cite{candes2004decode}, and $M\ge\rho K\log(N/K)$,%
\footnote{The parameter $\rho\in(0,1)$ is independent of $M$, $N$, and $K$.
An exact expression can be found in \cite{davenport2011introduction}.%
} we can recover $\boldsymbol{x}$ from $\boldsymbol{y}$ by considering
the following $\ell_{1}$-norm minimization problem,\begin{subequations}\label{eq:bpdn}
\begin{align}
\mathrm{minimize\;\;} & \|\boldsymbol{\theta}\|_{1}\\
\mathrm{subject}\;\mathrm{to\;\;} & \|\boldsymbol{A}\boldsymbol{T}^{-1}\boldsymbol{\theta}-\boldsymbol{y}\|_{2}\leq\varepsilon,\label{eq:bpcon-1}
\end{align}
\end{subequations}called BPDN \cite{candes2005stable}. If, for a
given matrix $\boldsymbol{A}$, there exists a constant $\delta_{K}\in(0,1)$
such that the following inequality holds for all $K$-sparse vectors
$\boldsymbol{v}$,
\begin{equation}
(1-\delta_{K})\|\boldsymbol{v}\|_{2}^{2}\leq\|\boldsymbol{A}\boldsymbol{v}\|_{2}^{2}\leq(1+\delta_{K})\|\boldsymbol{v}\|_{2}^{2},
\end{equation}
then $\boldsymbol{A}$ satisfies the RIP of order $K$. The computation
of $\delta_{K}$ is NP-hard. In \cite{candes2004decode} and \cite{baraniuk2007simple}
it was shown that if $\boldsymbol{A}$ is a Gaussian random matrix
with i.i.d. entries $A_{ml}\stackrel{}{\sim}\mathcal{N}(0,1/M)$,
then $\boldsymbol{A}$ satisfies the RIP with very high probability.

Assuming $\delta_{2K}<\sqrt{2}-1$ and $\varepsilon\ge\|\boldsymbol{e}_{\mathrm{obs}}\|_{2}$,
the $\ell_{2}$-norm of the reconstruction error of BPDN is upper
bounded by \cite{candes2005stable,candes2008rip}
\begin{equation}
\|\boldsymbol{x}-\boldsymbol{x}^{\star}\|_{2}\leq\frac{C_{0}}{\sqrt{K}}\|\boldsymbol{T}^{-1}(\boldsymbol{\theta}-\boldsymbol{\theta}_{K})\|_{1}+C_{1}\varepsilon,\label{eq:bound1-1}
\end{equation}
where $\boldsymbol{x}^{\star}=\boldsymbol{T}^{-1}\boldsymbol{\theta}^{\star}$,
in which $\boldsymbol{\theta}^{\star}$ is the solution to (\ref{eq:bpdn}),
$\boldsymbol{\theta}_{K}$ is the best $K$-sparse approximation of
the transformed underlying signal, and $C_{0},C_{1}\geq0$ are constants
that depend on $\delta_{K}$ \cite{candes2005stable}. Here, we only
consider strictly $K$-sparse signals, meaning there are at most $K$
non-zero components in $\boldsymbol{\theta}$. Hence, the first term
on the right hand side of (\ref{eq:bound1-1}) is zero.

Since the entries of $\boldsymbol{A}$ are i.i.d. Gaussian, \textcolor{black}{it
follows that} $\boldsymbol{e}_{\mathrm{obs}}\sim\mathcal{N}\left(0,(N/M)\sigma_{n}^{2}\boldsymbol{I}_{M}\right)$,
\textcolor{black}{so that} $\|\boldsymbol{e}_{\mathrm{obs}}\|_{2}$
is distributed according to a scaled $\chi_{M}$-distribution. Hence,
by Taylor series expansion, $\mathbb{E}_{\boldsymbol{A},\boldsymbol{n}}\left\{ \|\boldsymbol{e}_{\mathrm{obs}}\|_{2}\right\} \approx\sigma_{n}\sqrt{N}(1-1/(4M))$
and $\mathrm{Var}_{\boldsymbol{A},\boldsymbol{n}}(\|\boldsymbol{e}_{\mathrm{obs}}\|_{2})\approx(N/M)\sigma_{n}^{2}(1/2-1/(8M))$.
Therefore, to satisfy $\varepsilon\ge\|\boldsymbol{e}_{\mathrm{obs}}\|_{2}$
with high probability, $\varepsilon$ should be choosen as
\begin{align}
\varepsilon & =\varepsilon_{\mathrm{ref}}=\sigma_{n}\sqrt{N}\left(\left(1-\frac{1}{4M}\right)+\lambda\sqrt{\frac{1}{2M}-\frac{1}{8M^{2}}}\right)\label{eq:epsilon}\\
 & \approx\sigma_{n}\sqrt{N}\left(1+\lambda\sqrt{\frac{1}{2M}}\right),
\end{align}
where $\lambda\ge0$ is used to achieve a desired confidence level.

\subsection{Distributed Compressed Sensing for Networked Data}

We observe that the compression in (\ref{eq:compress2}) can be written
as a sum of linear projections of the measurements $z_{k}$ onto the
corresponding column $\boldsymbol{a}_{k}$ of $\boldsymbol{A}$,
\begin{equation}
\boldsymbol{y}=\boldsymbol{Az}=\sum_{k=0}^{N-1}\boldsymbol{a}_{k}z_{k}.\label{eq:lincombi}
\end{equation}
If we generate $\boldsymbol{a}_{k}$ in sensor $k$, it can compute
its contribution $\boldsymbol{w}_{k}=\boldsymbol{a}_{k}z_{k}$ to
the compression. By distributing the local projections $\boldsymbol{w}_{k}$
in the network using sensor-to-sensor communication and local processing
in the sensors, we can compute (\ref{eq:lincombi}) in a decentralized
manner. Consequently, this compression reduces the number of sensors
that need to convey information to the sink, effectively reducing
the communication cost at the expense of a delay induced by the local
communication. In Sections \ref{sec:DLPclu} and \ref{sec:DLPcon}
we present two approaches to such distributed processing for which
we determine the node-to-sink communication cost, inter-node communication
delay, and an upper bound on the reconstruction error.

\section{Distributed Linear Projections using Clustering\label{sec:DLPclu}}

\subsection{Cluster Formation and Operation\label{sub:clustersub}}

A set of nodes $\mathcal{L}\subseteq\mathcal{V}$, $|\mathcal{L}|=L$,
is selected to act as aggregating nodes (clusterheads), such that
clusterhead $i\in\mathcal{L}$ collects information from a subset
(cluster) $\mathcal{C}_{i}\subseteq\mathcal{V}$, of the sensors in
the network. The clusterhead selection is done with respect to the
local communication range such that each clusterhead is located at
the center of its cluster, which has radius $R$. The clusters are
disjoint, i.e., $\mathcal{C}_{i}\cap\mathcal{C}_{l}=\emptyset$ for
$i\neq l$, and $\bigcup_{i\in\mathcal{L}}\mathcal{C}_{i}=\mathcal{V}$.
Note that depending on $N$ and $R$, one of the clusters at the boundary
may be smaller than the others. The number of clusters is given by
\begin{equation}
L=\left\lceil \frac{N}{2R+1}\right\rceil .\label{eq:clsize}
\end{equation}

Node $k$ computes its local linear projection $\boldsymbol{w}_{k}=\boldsymbol{a}_{k}z_{k}$
and sends it to its clusterhead. Clusterhead $i$ computes
\begin{equation}
\boldsymbol{y}_{\mathcal{C}_{i}}=\sum_{k\in\mathcal{C}_{i}}\boldsymbol{w}_{k}.
\end{equation}
Finally, the clusterheads transmit their partial information to the
sink. Since the clusters are disjoint, the sink computes $\boldsymbol{y}=\boldsymbol{y}_{\mathcal{C}_{1}}+\dots+\boldsymbol{y}_{\mathcal{C}_{L}}=\boldsymbol{Az}$,
and reconstructs $\boldsymbol{x}$ using BPDN.

\subsection{Cost and Delay}

The total communication cost is $C_{\mathrm{tot}}=CL$. The delay
is given by the number of time slots in the S-TDMA needed to schedule
a broadcast transmission for every non-clusterhead node. Due to the
cluster formation and communication model that we consider, there
is no interference from nodes in a cluster to the neighboring clusterheads.
Hence, the delay $D$ is given by the maximum node degree of the clusterheads,
\begin{equation}
D=\begin{cases}
2R & \quad0\leq2R<N\\
N-1 & \quad2R\ge N.
\end{cases}\label{eq:cldelay}
\end{equation}

\subsection{Reconstruction Performance and Robustness}

Define the set $\mathcal{D}\subseteq\mathcal{L}$ as the set of clusterheads
whose packets are erased during transmission to the sink. The sink
is assumed to have no knowledge of $\mathcal{D}$, but attempts to
recover $\boldsymbol{x}$ assuming it has received the correct compression
$\boldsymbol{y}=\boldsymbol{Az}$. The resulting compression at the
sink given a set of packet erasures $\mathcal{D}$ is 
\begin{align}
\tilde{\boldsymbol{y}} & =\sum_{i\in\mathcal{L}\setminus\mathcal{D}}\boldsymbol{y}_{\mathcal{C}_{i}}=\boldsymbol{Ax}+(\boldsymbol{A}-\boldsymbol{B})\boldsymbol{n}-\boldsymbol{Bx}=\boldsymbol{y}-\boldsymbol{Bz},\label{eq:clustdrop}
\end{align}
where $\boldsymbol{B}$ is a matrix \textcolor{black}{whose nonzero}
columns\textcolor{black}{,}\textcolor{blue}{{} }\textcolor{black}{corresponding}
to the nodes whose clusterhead packet was erased\textcolor{blue}{,
}\textcolor{black}{are} equal to the corresponding columns of $\boldsymbol{A}$.
Therefore, for packet erasure probability $p=0$, we have $\boldsymbol{B}=\boldsymbol{0}$
and $\tilde{\boldsymbol{y}}=\boldsymbol{y}$, while for $p\neq0$
we have to account for $\boldsymbol{Bz}$ when setting $\varepsilon$
in the BPDN for (\ref{eq:bound1-1}) to hold. The following Theorem
\textcolor{black}{describes how $\varepsilon$ should be selected.}
\begin{thm}
\label{thm:clustthm}Given the model described in Section \ref{sec:SM},
$\boldsymbol{A}$ as described in Section \ref{sub:defper}, and the
compression in (\ref{eq:clustdrop}),  the choice of $\varepsilon$
that guarantees a stable recovery using BPDN is
\begin{equation}
\varepsilon=\varepsilon_{\mathrm{ref}}\sqrt{1-\left(1-\frac{1-p}{1-p^{\left\lceil N/(2R+1)\right\rceil }}\right)\left(1-\mathsf{SNR}\right)}.\label{eq:epsclust}
\end{equation}

\end{thm}
The proof is given in Appendix \ref{sec:proofclust}.

\section{Distributed Linear Projections using Consensus\label{sec:DLPcon}}

An alternative approach is to compute $\boldsymbol{y}$ from (\ref{eq:lincombi})
directly in the network by using a fully distributed algorithm. Here,
we propose the use of average consensus.

\subsection{The Consensus Algorithm}

We can express (\ref{eq:lincombi}) as
\begin{equation}
\boldsymbol{y}=\sum_{k=0}^{N-1}\boldsymbol{w}_{k}=N\left(\frac{1}{N}\sum_{k=0}^{N-1}\boldsymbol{w}_{k}\right)=N\bar{\boldsymbol{w}}.\label{eq:avgz}
\end{equation}
We use average consensus to estimate $\bar{\boldsymbol{w}}$ in the
network. The estimate is then used at the sink to compute (\ref{eq:avgz}).
Let $\boldsymbol{w}_{k}(0)=\boldsymbol{w}_{k}$ be the initial value
at sensor $k$. The updating rule of average consensus \cite{olfati2007con}
is given by
\begin{equation}
\boldsymbol{w}_{k}(i)=\boldsymbol{w}_{k}(i-1)+\xi\sum_{v\in\mathcal{M}_{k}}\left(\boldsymbol{w}_{k}(i-1)-\boldsymbol{w}_{v}(i-1)\right),\label{eq:consensus}
\end{equation}
where $\mathcal{M}_{k}\subset\mathcal{V}$ is the set of neighboring
sensors of sensor $k$, $\xi$ is the algorithm step size, and $i$
is the iteration index. We can also express (\ref{eq:consensus})
in matrix form as
\begin{equation}
\boldsymbol{W}(i)=\boldsymbol{P}\boldsymbol{W}(i-1)=\boldsymbol{P}^{i}\boldsymbol{W}(0),
\end{equation}
where $\boldsymbol{W}(0)=[\boldsymbol{w}_{1}(0),\dots,\boldsymbol{w}_{N}(0)]^{\mathsf{T}}$
and $\boldsymbol{P}=\boldsymbol{I}_{N}-\xi\boldsymbol{L}$, in which
$\boldsymbol{L}$ denotes the graph Laplacian of $\mathcal{G}$. By
properties of the consensus algorithm \cite{olfati2007con}, $\bar{\boldsymbol{w}}$
is conserved in each iteration,
\begin{equation}
\bar{\boldsymbol{w}}=\frac{1}{N}\sum_{k=0}^{N-1}\boldsymbol{w}_{k}(0)=\frac{1}{N}\sum_{k=0}^{N-1}\boldsymbol{w}_{k}(i),
\end{equation}
irrespective of $i$. If $\xi$ is chosen small enough \cite{olfati2007con},
the algorithm is monotonically converging in the limit $i\rightarrow+\infty$
to the average in all sensor nodes, 
\begin{equation}
\lim_{i\rightarrow\infty}\boldsymbol{w}_{k}(i)=\bar{\boldsymbol{w}}.
\end{equation}

After a certain number of iterations $I$, a set $\mathcal{L}\subseteq\mathcal{V}$,
with $|\mathcal{L}|=L$, of randomly chosen sensors communicate their
estimates $\boldsymbol{w}_{k}(I)$ of $\bar{\boldsymbol{w}}$ to the
sink. However, due to erasures, the sink estimates $\boldsymbol{y}$
from a set of nonerased packets $\tilde{\mathcal{L}}\subseteq\mathcal{L}$,
$|\tilde{\mathcal{L}}|=\tilde{L}$, as
\begin{equation}
\hat{\boldsymbol{y}}=N\left(\frac{1}{\tilde{L}}\sum_{k\in\mathcal{\tilde{L}}}\boldsymbol{w}_{k}(I)\right).\label{eq:conest}
\end{equation}
Finally, $\hat{\boldsymbol{y}}$ (cf. (\ref{eq:avgz})) is used in
(\ref{eq:bpdn}) to reconstruct $\boldsymbol{x}$.

\subsection{Cost and Delay}

As for clustering, the total communication cost is $C_{\mathrm{tot}}=CL$.
The delay is given by $D_{\mathrm{cons}}=DI$\textcolor{black}{, where
$D$ was defined in (\ref{eq:cldelay}).}

\subsection{Reconstruction Performance and Robustness}

Due to the fact that average consensus only converges in the limit
$i\rightarrow+\infty$, for any finite $I$ there will be an error
in each sensor estimate $\boldsymbol{w}_{k}(I)$ with respect to the
true average $\bar{\boldsymbol{w}}$. The transmitted packets to the
sink can also be erased. This results in a mismatch $\boldsymbol{e}_{\mathrm{cons}}=\boldsymbol{y}-\hat{\boldsymbol{y}}$
between the desired compression and the compression calculated using
average consensus. As for the clustering case, in order to guarantee
that the reconstruction error is upper bounded by (\ref{eq:bound1-1}),
this perturbation has to be accounted for when setting $\varepsilon$
in the BPDN. The following Theorem states how this should be done.
\begin{thm}
\label{thm:consthm}Let $\mu_{2}$ be the second largest eigenvalue
of $\boldsymbol{P}$. Given the model in Section \ref{sec:SM}, $\boldsymbol{A}$
as described in Section \ref{sub:defper}, and the compression in
(\ref{eq:conest}),  the choice of $\varepsilon$ that guarantees
a stable recovery using BPDN is
\begin{equation}
\varepsilon=\varepsilon_{\mathrm{ref}}\left(1+\mu_{2}^{I}\sqrt{\left(1+\mathsf{SNR}\right)\Phi}\right),\label{eq:epscon}
\end{equation}
where%
\footnote{The value of $\varepsilon$ in (\ref{eq:epscon}) may be very conservative,
since the upper bound on the convergence rate of consensus (see (\ref{eq:varep})
in Appendix \ref{sec:proofcons}) may be very loose. Consider the
eigenvalue decomposition of $\boldsymbol{P}=\boldsymbol{Q}\boldsymbol{M}\boldsymbol{Q}^{-1}$,
and let $\boldsymbol{\alpha}=\boldsymbol{Q}^{-1}\boldsymbol{w}$ be
the projection of the data $\boldsymbol{w}$ onto the eigenspace of
$\boldsymbol{P}$. Order the eigenvalues \textcolor{black}{of $\bm{P}$}
as $1=\mu_{1}>\mu_{2}\ge\dots\ge\mu_{N}$. Then, the disagreement
after $I$ iterations of consensus on $\boldsymbol{w}$ is given by\setcounter{equation}{22}
\begin{align}
\|\boldsymbol{w}(I)-\bar{\boldsymbol{w}}\|_{2}^{2} & =\sum_{k=2}^{N}\mu_{k}^{2I}\alpha_{k}^{2},\label{eq:exacterr}
\end{align}
which can be upper bounded by
\begin{equation}
\sum_{k=2}^{N}\mu_{k}^{2I}\alpha_{k}^{2}\le\mu_{2}^{2I}\sum_{k=2}^{N}\alpha_{k}^{2}\overset{(\mathrm{a})}{=}\mu_{2}^{2I}\|\boldsymbol{w}(0)-\bar{\boldsymbol{w}}\|_{2}^{2},\label{eq:upperberr}
\end{equation}
where $(\mathrm{a})$ follows since $\alpha_{1}$ is the entry that
corresponds to the eigenvector of $\mu_{1}$, and thus the initial
disagreement is $\|\boldsymbol{w}(0)-\bar{\boldsymbol{w}}\|_{2}^{2}=\sum_{k=2}^{N}\alpha_{k}^{2}$.
In general, (\ref{eq:exacterr}) is hard to compute and (\ref{eq:upperberr})
may be loose. If the support of $\boldsymbol{\alpha}$ is concentrated
to those entries corresponding to the smaller eigenvalues, $\|\boldsymbol{w}(I)-\bar{\boldsymbol{w}}\|_{2}^{2}$
decreases much faster than $\mu_{2}^{2I}$ in the first iterations.
However, after enough iterations the smaller eigenvalues have diminished,
and the convergence rate is dominated by $\mu_{2}$. In our case,
the data $\boldsymbol{w}$ is Gaussian, and since the columns of $\bm{Q}^{-1}$
form an orthonormal basis in $\mathbb{R}^{N}$, $\bm{\alpha}$ is
also Gaussian with the same mean and variance. Therefore, the power
of $\bm{\alpha}$ is spread evenly in its entries. Consequently, the
bound is loose for our signals and consensus behaves much better with
respect to $I$ than shown in Figs. \ref{fig:codel} and \ref{fig:robust}.%
}\setcounter{equation}{21} 
\begin{equation}
\Phi=\frac{N(1-p{}^{L})}{L(1-p)}\left(\frac{N-\frac{L(1-p)}{(1-p^{L})}}{N-1}\right).
\end{equation}

\end{thm}
The proof is given in Appendix \ref{sec:proofcons}.

\section{Results and Discussion}

In this section, we evaluate the cost-delay tradeoff of the clustering
and consensus approaches, i.e., how the reconstruction error scales
with the number of iterations $I$ and the number of nodes transmitting
to the sink $L$, and compare the robustness to packet erasures. We
fix $N=100$, $M=20$, $R=10$, $\sigma_{n}^{2}=0.01$, and $E_{X}=3$,
giving $\mathsf{SNR}=3$ in linear scale. The figures are created
by computing $\varepsilon$ using the expressions in Theorems \ref{thm:clustthm}
and \ref{thm:consthm}, where the upper bound \textcolor{black}{$\mu_{2}\le\cos(\pi R/2N)$
}is used in (\ref{eq:epscon}), assuming $\xi$ is chosen optimally
\cite{olfati2007con}. Since $C_{1}$ in (\ref{eq:bound1-1}) is NP-hard
to compute, we normalize the error with respect to $C_{1}$. Also,
since $M$, $N$, and $\sigma_{n}^{2}$ are fixed, we also normalize
with respect to $\varepsilon_{\mathrm{ref}}$. Hence, the normalized
error is equal to $\zeta_{\mathrm{norm}}=\|\boldsymbol{x}-\boldsymbol{x}^{\star}\|_{2}/(C_{1}\varepsilon_{\mathrm{ref}})$.
Note that $\zeta_{\mathrm{norm}}\ge1$.

\subsection{Cost-Delay Tradeoff}

\begin{figure}[t]
\psfrag{xlabel}{\footnotesize{$C_{\mathrm{tot}}$}}\psfrag{ylabel}{\footnotesize{$D_{\mathrm{cons}}$}}\psfrag{legend1legend}{\scriptsize{$p=0$}}\psfrag{legend2legend}{\scriptsize{$p=0.05$}}\psfrag{legend3legend}{\scriptsize{$p=0.2$}}\psfrag{legend4}{\scriptsize{$p=0.5$}}\includegraphics[width=1\columnwidth]{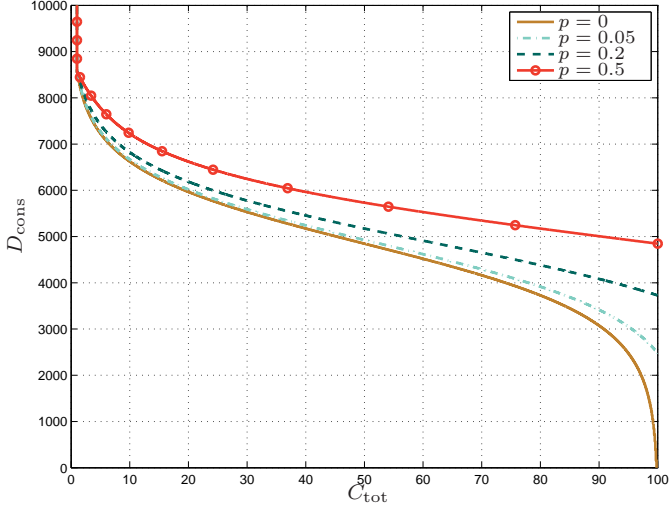}\caption{\label{fig:codel}The figure shows the boundaries of systems satisfying
the given error threshold $\nu$. The area above the graphs are the
regions of points $(C_{\mathrm{tot}},D_{\mathrm{cons}})$ satisfying
$\zeta_{\mathrm{norm}}\le\nu$, where $\nu=1.1$, for $p\in\{0,0.05,0.2,0.5\}$,
when using average consensus with $R=10$. }
\end{figure}
Fig. \textcolor{black}{\ref{fig:codel}} shows the boundaries of the
regions giving a normalized error lower than the threshold\textcolor{black}{{}
$\nu=1.1$} for packet erasure probabilities $p\in\{0,0.05,0.2,0.5\}$.
As can be seen, a higher packet erasure probability results in a boundary
receding towards the top right corner, meaning that higher $I$ and
$L$ are needed to meet $\nu$. An important observation is also that
the normalized error $\zeta_{\mathrm{norm}}$ is nonincreasing in
$I$ and $L$. Looking at the slope of the curves, we see that there
are differences in how much delay we must \textcolor{black}{tolerate}
in order to lower communication cost. For example when $p=0$, for
low and high costs, we need to increase delay significantly, while
for medium costs the curves are flatter and a smaller increase in
delay is \textcolor{black}{sufficient} to reduce cost.

For clustering, $D$ and $C_{\mathrm{tot}}$ are implicitly given
by $R$ through (\ref{eq:cldelay}) and (\ref{eq:clsize}). Hence,
there is no tradeoff as such for the clustering. The implication is
that for larger $p$ we cannot increase cost or delay to ensure that
$\zeta_{\mathrm{norm}}\le\nu$.

\subsection{Robustness to Packet Erasures}

\begin{figure}
\psfrag{xlp}{\footnotesize{$p$}}\psfrag{ylabel}{\footnotesize{$\zeta_{\mathrm{norm}}$}}\psfrag{legend1}{\scriptsize{clustering}}\psfrag{legend2legendlegend}{\scriptsize{consensus, $I=400$}}\psfrag{legend3}{\scriptsize{consensus, $I=300$}}\includegraphics[width=1\columnwidth]{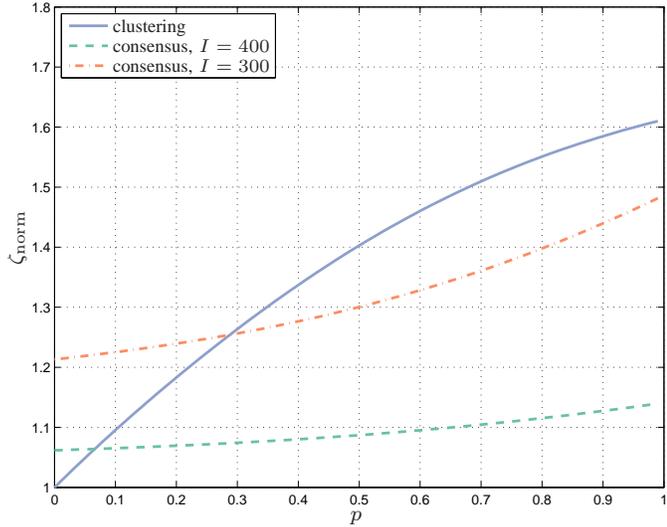}

\caption{\label{fig:robust}This plot shows $\zeta_{\mathrm{norm}}$ for different
packet erasure probabilities of clustering and two cases of consensus
with $I\in\{300,400\}$, where $R=10$ and $L=\left\lceil N/(2R+1)\right\rceil $}
\end{figure}
Fig. \textcolor{black}{\ref{fig:robust}} depicts the behavior of
$\zeta_{\mathrm{norm}}$ with respect to $p$. \textcolor{black}{From
the slope of the curves we see that} consensus is less sensitive to
packet erasures as compared to clustering. This is in line with the
results in Theorems \ref{thm:clustthm} and \ref{thm:consthm}, where
$\zeta_{\mathrm{norm}}\propto\sqrt{p}$ and $\zeta_{\mathrm{norm}}\propto1/\sqrt{1-p}$,
for clustering and consensus, respectively (see (\ref{eq:epsclust})
and (\ref{eq:epscon})). Note that the source of error is different
for clustering and consensus. Both approaches are affected by packet
erasures, but in different manners. For clustering, if an erasure
occurs, that information is lost, while for consensus the estimation
step (\ref{eq:conest}) at the sink is affected only to a small degree.
This is because the consensus algorithm disseminates the information
throughout the network, making it more robust to packet erasures.
On the other hand, for consensus $\zeta_{\mathrm{norm}}$ is dominated
by the disagreement between the estimates at the nodes and the true
average. This explains the superiority of clustering for small $p$.
However, the disagreement decreases exponentially in $I$, so $\zeta_{\mathrm{norm}}$
can be made arbitrarily small by increasing $I$.

\section{Conclusion}

We derived closed-form expressions for the upper bound on the $\ell_{2}$-norm
of the reconstruction error for a clustering and a consensus approach
to distributed compressed sensing in WSNs. For the consensus approach,
the expression can be used to trade off cost and delay such that the
reconstruction error is guaranteed to satisfy a given performance
requirement with high probability. We also analyzed the robustness
to erasures of packets sent to the sink. If a large enough number
of iterations is allowed, consensus is more robust than clustering,
except for very small packet erasure probabilities. Moreover, by increasing
the number of iterations, the additional error caused by the consensus
algorithm and packet erasures can be made arbitrarily small. Another
benefit of the consensus is that there is no need to \textcolor{black}{form
clusters}, which can be a hard task, especially if the sensors are
mobile. Future research includes unreliabe sensor-to-sensor communication,
uncertainty in the position of the nodes, \textcolor{black}{and more
general network topologies.}

\appendices{}

\section{Proof of Theorem 1\label{sec:proofclust}}

When using clustering, the compressed vector received by the sink
is\setcounter{equation}{24}
\begin{equation}
\tilde{\boldsymbol{y}}=\boldsymbol{Ax}+(\boldsymbol{A}-\boldsymbol{B})\boldsymbol{n}-\boldsymbol{Bx}.
\end{equation}
Define $\boldsymbol{C}=\boldsymbol{A}-\boldsymbol{B}$, and the total
perturbation $\boldsymbol{u}=\boldsymbol{Cn}-\boldsymbol{Bx}$. Let
$\boldsymbol{c}_{k}$ and $\boldsymbol{b}_{k}$ be the $k$th column
vector of $\boldsymbol{C}$ and $\boldsymbol{B}$, respectively. Then,
for each node $k$ we have $\boldsymbol{u}_{k}=\boldsymbol{c}_{k}n_{k}-\boldsymbol{b}_{k}x_{k}$,
and $\boldsymbol{u}=\sum_{k=0}^{N-1}\boldsymbol{u}_{k}$. Therefore,
for (\ref{eq:bound1-1}) to hold, we need $\varepsilon\ge\|\boldsymbol{u}\|_{2}$.
Denote by $\mathcal{H}$ the set of nodes whose information is not
erased, i.e., $\mathcal{H}=\{k\;:\; k\in\mathcal{C}_{j},j\notin\mathcal{D}\}$,
where $|\mathcal{H}|=H$. Note that $H$ is a zero-truncated binomial
random variable with parameters $L$ and $p$. It is easy to see that
$\mathbb{E}_{\mathcal{H},\boldsymbol{A},\boldsymbol{n}}\left\{ \boldsymbol{u}_{k}\right\} =\boldsymbol{0}$,
hence the covariance matrix is
\begin{align}
\mathbb{E}_{\mathcal{H},\boldsymbol{A},\boldsymbol{n}}\left\{ \boldsymbol{u}_{k}\boldsymbol{u}_{k}^{\mathsf{T}}\right\}  & =\mathbb{E}_{\mathcal{H},\boldsymbol{A}}\left\{ \boldsymbol{c}_{k}\boldsymbol{c}_{k}^{\mathsf{T}}\right\} \mathbb{E}_{\boldsymbol{n}}\left\{ n_{k}^{2}\right\} \nonumber \\
 & +\mathbb{E}_{\mathcal{H},\boldsymbol{A}}\left\{ \boldsymbol{b}_{k}\boldsymbol{b}_{k}^{\mathsf{T}}\right\} x_{k}^{2}.
\end{align}
For notational convenience, we drop the subscript indicating over
which variable the expectation is taken. We observe that for $k\notin\mathcal{H}$,
$\boldsymbol{c}_{k}=\boldsymbol{0}$ and $\boldsymbol{b}_{k}=\boldsymbol{a}_{k}$,
and for $k\in\mathcal{H}$, $\boldsymbol{c}_{k}=\boldsymbol{a}_{k}$
and $\boldsymbol{b}_{k}=\boldsymbol{0}$, thus
\begin{equation}
\mathbb{E}\left\{ \boldsymbol{u}_{k}\boldsymbol{u}_{k}^{\mathsf{T}}\right\} =\begin{cases}
\mathbb{E}\left\{ \boldsymbol{c}_{k}\boldsymbol{c}_{k}^{\mathsf{T}}\right\} \mathbb{E}\left\{ n_{k}^{2}\right\} =\frac{\sigma_{n}^{2}}{M}\boldsymbol{I}_{M} & \quad k\in\mathcal{H}\\
\mathbb{E}\left\{ \boldsymbol{b}_{k}\boldsymbol{b}_{k}^{\mathsf{T}}\right\} x_{k}^{2}=\frac{x_{k}^{2}}{M}\boldsymbol{I}_{M} & \quad k\notin\mathcal{H}.
\end{cases}
\end{equation}
It follows that
\begin{align}
\mathbb{E}\left\{ \boldsymbol{u}\boldsymbol{u}^{\mathsf{T}}\right\}  & =\mathbb{E}\left\{ \left(\sum_{k=0}^{N-1}\boldsymbol{u}_{k}\right)\left(\sum_{k=0}^{N-1}\boldsymbol{u}_{k}\right)^{\mathsf{T}}\right\} \\
 & \overset{(\mathrm{a})}{=}\mathbb{E}\left\{ \sum_{k=0}^{N-1}\boldsymbol{u}_{k}\boldsymbol{u}_{k}^{\mathsf{T}}\right\} \\
 & =\sum_{k=0}^{N-1}\left(\mathbb{E}\left\{ \boldsymbol{u}_{k}\boldsymbol{u}_{k}^{\mathsf{T}}\mathbb{I}_{\mathcal{H}}(k)\right\} +\mathbb{E}\left\{ \boldsymbol{u}_{k}\boldsymbol{u}_{k}^{\mathsf{T}}\mathbb{I}_{\mathcal{V}\setminus\mathcal{H}}(k)\right\} \right),
\end{align}
where $(\mathrm{a})$ follows since all $\boldsymbol{u}_{k}$'s are
mutually independent. The probability that $k\notin\mathcal{H}$ is
given by
\begin{equation}
p_{\mathcal{H}}=1-\frac{1-p}{1-p^{\left\lceil N/(2R+1)\right\rceil }}.
\end{equation}
Then, we have
\begin{align}
\mathbb{E}\left\{ \boldsymbol{u}\boldsymbol{u}^{\mathsf{T}}\right\}  & =\frac{1}{M}\left(N(1-p_{\mathcal{H}})\sigma_{n}^{2}+Np_{\mathcal{H}}\frac{E_{X}}{N}\right)\boldsymbol{I}_{M}\\
 & =\frac{\sigma_{n}^{2}N}{M}\left(1-p_{\mathcal{H}}(1-\mathsf{SNR})\right)\boldsymbol{I}_{M}\triangleq\sigma_{\boldsymbol{u}}^{2}\boldsymbol{I}_{M}.\label{eq:sigu}
\end{align}
For large enough $N$, $\boldsymbol{u}\sim\mathcal{N}(\boldsymbol{0},\sigma_{\boldsymbol{u}}^{2}\boldsymbol{I}_{M})$,
and consequently $\|\boldsymbol{u}\|_{2}$ is distributed according
to a scaled $\chi_{M}$-distribution. Hence, $\mathbb{E}\left\{ \|\boldsymbol{u}\|_{2}\right\} =\sigma_{\boldsymbol{u}}\sqrt{M}(1-1/4M)$
and $\mathrm{Var}\left(\|\boldsymbol{u}\|_{2}\right)=\sigma_{\boldsymbol{u}}^{2}(1/2-1/8M)$.
Therefore, using $\sigma_{\boldsymbol{u}}^{2}$ as defined in (\ref{eq:sigu}),
the robust choice for $\varepsilon$ is
\begin{align}
\varepsilon & =\mathbb{E}\left\{ \|\boldsymbol{u}\|_{2}\right\} +\lambda\sqrt{\mathrm{Var}\left(\|\boldsymbol{u}\|_{2}\right)}\label{eq:epsclu}\\
 & =\varepsilon_{\mathrm{ref}}\sqrt{1-\left(1-\frac{1-p}{1-p^{\left\lceil N/(2R+1)\right\rceil }}\right)\left(1-\mathsf{SNR}\right)}.
\end{align}

\section{Proof of Theorem 2\label{sec:proofcons}}

The vector received by the sink using consensus is
\begin{equation}
\hat{\boldsymbol{y}}=\boldsymbol{Ax}+\boldsymbol{e}_{\mathrm{obs}}+\boldsymbol{e}_{\mathrm{cons}}.
\end{equation}
In order to guarantee stable reconstruction $\varepsilon\ge\|\boldsymbol{e}_{\mathrm{obs}}+\boldsymbol{e}_{\mathrm{cons}}\|_{2}$.
By the triangle inequality, we have
\begin{equation}
\|\boldsymbol{e}_{\mathrm{obs}}+\boldsymbol{e}_{\mathrm{cons}}\|_{2}\le\|\boldsymbol{e}_{\mathrm{obs}}\|_{2}+\|\boldsymbol{e}_{\mathrm{cons}}\|_{2}.\label{eq:triineq}
\end{equation}
Thus, we choose $\varepsilon\ge\|\boldsymbol{e}_{\mathrm{obs}}\|_{2}+\|\boldsymbol{e}_{\mathrm{cons}}\|_{2}$.
The statistics of the first term on the right hand side of (\ref{eq:triineq})
are given in Section \ref{sec:csback}. It remains to determine the
contribution from the consensus. Since all dimensions of $\hat{\boldsymbol{y}}$
are i.i.d. we can calculate the statistics from one dimension and
deduce what the total contribution is. We fix the number of iterations
$I$, the number of queried nodes $L$, and the data \textcolor{black}{$\bm{w}_{k}$}.
Define the disagreement between the estimate from $\tilde{L}$ received
packets and the true average after $I$ iterations for each dimension
$m=1,\dots,M$ \textcolor{black}{as}
\begin{equation}
\Delta_{m}(I,L)=\frac{1}{\tilde{L}}\left(\sum_{k\in\mathcal{\tilde{L}}}w_{k,m}(I)\right)-\bar{w}_{m}=\hat{w}_{m}(I,\tilde{L})-\bar{w}_{m},
\end{equation}
where $w_{k,m}(I)$ is the $m$th element of the vector $\boldsymbol{w}_{k}(I)$,
$\bar{w}_{m}$ is the average over the $m$th dimension, and $\hat{w}_{m}(I,\tilde{L})$
is the estimate \textcolor{black}{of $\bar{w}_{m}$}. For notational
convenience we drop the subscript indicating the dimension, and the
dependencies on $I$ and $L$. Now, there are two sources of randomness:
(i) the set of queried nodes $\mathcal{L}\subseteq\mathcal{V}$, which
is randomly selected; (ii) the number of nonerased packets $\tilde{L}\le L$,
due to random packet erasures. Since $w_{k}$'s are fixed, $\bar{w}$
is constant. Hence,
\begin{align}
\mathrm{Var}_{\tilde{L},\mathcal{L}}(\Delta) & =\mathrm{Var}_{\tilde{L},\mathcal{L}}\left(\hat{w}\right)=\mathbb{E}_{\tilde{L},\mathcal{L}}\left\{ (\hat{w}-\bar{w})^{2}\right\} \\
 & =\mathbb{E}_{\tilde{L}}\left\{ \mathbb{E}_{\left.\mathcal{L}\right|\tilde{L}}\left\{ (\hat{w}-\bar{w})^{2}\right\} \right\} .
\end{align}
The estimate $\hat{w}$ is an estimate by simple random sampling from
a finite population of size $N$. Then
\begin{align}
\mathbb{E}_{\tilde{L}} & \left\{ \mathbb{E}_{\left.\mathcal{L}\right|\tilde{L}}\left\{ (\hat{w}-\bar{w})^{2}\right\} \right\} \\
 & =\mathbb{E}_{\tilde{L}}\left\{ \frac{1}{\tilde{L}N}\left(\sum_{k=0}^{N-1}(w_{k}-\bar{w})^{2}\right)\frac{N-\tilde{L}}{N-1}\right\} \\
 & =\frac{1}{N(N-1)}\left(\sum_{k=0}^{N-1}(w_{k}-\bar{w})^{2}\right)\left(N\mathbb{E}_{\tilde{L}}\left\{ \frac{1}{\tilde{L}}\right\} -1\right),
\end{align}
where the first equality is due to \cite[Thm. 7.3.1B]{rice2006mathematical}.
The expectation of the \textcolor{black}{inverse} of $\tilde{L}$
is \cite{rempala2004asymptotic}
\begin{equation}
\mathbb{E}_{\tilde{L}}\left\{ \frac{1}{\tilde{L}}\right\} \approx(1-p^{L})\frac{1}{L(1-p)}\triangleq\frac{1}{\bar{L}}.
\end{equation}
If we consider again the dependence of $w_{k}$ on $I$ and let $\boldsymbol{w}(I)=[w_{1}(I),\dots,w_{N}(I)]^{\mathsf{T}}$,
we have
\begin{equation}
\mathrm{Var}_{\tilde{L},\mathcal{L}}(\Delta)=\frac{1}{\bar{L}}\left(\frac{N-\bar{L}}{N(N-1)}\right)\|\boldsymbol{w}(I)-\bar{w}\boldsymbol{1}\|_{2}^{2}.
\end{equation}
The convergence rate of the $\ell_{2}$-norm $\|\boldsymbol{w}(I)-\bar{w}\boldsymbol{1}\|_{2}$
is defined as
\begin{equation}
\varrho=\lim_{I\rightarrow\infty}\left(\frac{\|\boldsymbol{w}(I)-\bar{w}\boldsymbol{1}\|_{2}}{\|\boldsymbol{w}(0)-\bar{w}\boldsymbol{1}\|_{2}}\right)^{1/I}\;,\quad\boldsymbol{w}(0)\neq\bar{w}\boldsymbol{1},
\end{equation}
which can be upper bounded by $\varrho\le\mu_{2}^{I}$ \cite{olfati2007con}.
Consequently, we have
\begin{equation}
\|\boldsymbol{w}(I)-\bar{w}\boldsymbol{1}\|_{2}^{2}\le\|\boldsymbol{w}(0)-\bar{w}\boldsymbol{1}\|_{2}^{2}\mu_{2}^{2I}.\label{eq:varep}
\end{equation}
Now, considering the randomness of $\boldsymbol{A}$ and $\boldsymbol{n}$,
\begin{equation}
\mathrm{Var}_{\tilde{L},\mathcal{L},\boldsymbol{A},\boldsymbol{n}}(\Delta)\le\frac{1}{\bar{L}}\left(\frac{N-\bar{L}}{N(N-1)}\right)\mathbb{E}_{\boldsymbol{A},\boldsymbol{n}}\left\{ \|\boldsymbol{w}(0)-\bar{w}\boldsymbol{1}\|_{2}^{2}\right\} \mu_{2}^{2I}.
\end{equation}
Furthermore
\begin{align}
\mathbb{E}_{\boldsymbol{A},\boldsymbol{n}}\left\{ \|\boldsymbol{w}(0)-\bar{w}\boldsymbol{1}\|_{2}^{2}\right\}  & =\sum_{k=0}^{N-1}\mathbb{E}_{\boldsymbol{A},\boldsymbol{n}}\left\{ (w_{k}(0)-\bar{w})^{2}\right\} \\
 & =\frac{N-1}{N}\left(\sum_{k=0}^{N-1}\frac{\sigma_{n}^{2}}{M}+\sum_{k=0}^{N-1}\frac{x_{k}^{2}}{M}\right)\\
 & \approx\frac{N\sigma_{n}^{2}+E_{X}}{M},
\end{align}
where the last step follows since we consider very large $N$. Finally,
we have
\begin{equation}
\sigma_{\Delta}^{2}\triangleq\mathrm{Var}_{\tilde{L},\mathcal{L},\boldsymbol{A},\boldsymbol{n}}(\Delta)\le\frac{1}{\bar{L}}\left(\frac{N-\bar{L}}{N-1}\right)\frac{E_{X}+N\sigma_{n}^{2}}{NM}\mu_{2}^{2I}.
\end{equation}
Due to the multiplication by $N$ in (\ref{eq:conest}), and since
all dimensions $m$ are i.i.d., $\boldsymbol{e}_{\mathrm{cons}}\sim\mathcal{N}(0,N^{2}\sigma_{\Delta}^{2}\boldsymbol{I}_{M})$,
hence $\|\boldsymbol{e}_{\mathrm{cons}}\|_{2}$ is distributed according
to a scaled $\chi_{M}$-distribution with $\mathbb{E}\left\{ \|\boldsymbol{e}_{\mathrm{cons}}\|_{2}\right\} =N\sigma_{\Delta}\sqrt{M}(1-1/4M)$
and $\mathrm{Var}\left(\|\boldsymbol{e}_{\mathrm{cons}}\|_{2}\right)=N^{2}\sigma_{\Delta}^{2}(1/2-1/8M)$.
Using (\ref{eq:triineq}) and the same argument as in (\ref{eq:epsclu}),
the robust choice of $\varepsilon$ is
\begin{equation}
\varepsilon=\varepsilon_{\mathrm{ref}}\left(1+\mu_{2}^{I}\sqrt{\left(1+\mathsf{SNR}\right)\Phi}\right),
\end{equation}
where
\begin{equation}
\Phi=\frac{N(1-p{}^{L})}{L(1-p)}\left(\frac{N-\frac{L(1-p)}{(1-p^{L})}}{N-1}\right).
\end{equation}

\section*{}

\section*{}

\bibliographystyle{IEEEtran}
\bibliography{BibCS}

\end{document}